\documentclass[reprint,aps,pre,floatfix,superscriptaddress,showpacs,amsmath,amssymb]{revtex4-2}

\usepackage{graphicx}% Include figure files
\usepackage{dcolumn}% Align table columns on decimal point
\usepackage{subfigure}% for two column figures
\usepackage{bm}% bold math
\usepackage{color}
\usepackage[colorlinks,citecolor=red,urlcolor=blue,bookmarks=false,hypertexnames=true]{hyperref}

\newcommand{\bmath}{\begin{displaymath}}
\newcommand{\emath}{\end{displaymath}}

\newcommand{\be}{\begin{equation}}
\newcommand{\ee}{\end{equation}}
\newcommand{\bea}{\begin{eqnarray}}
\newcommand{\eea}{\end{eqnarray}}

\newcommand{\bmultl}{\begin{multline}}
\newcommand{\emultl}{\end{multline}}

\newcommand{\bsubeq}{\begin{subequations}}
\newcommand{\esubeq}{\end{subequations}}
\newcommand{\bitemize}{\begin{itemize}}
\newcommand{\eitemize}{\end{itemize}}

\newcommand{\bmx}{\begin{bmatrix}}
\newcommand{\emx}{\end{bmatrix}}
\newcommand{\bsmx}{\begin{smallmatrix}}
\newcommand{\esmx}{\end{smallmatrix}}

\begin{document}

\title{Diffraction and pseudospectra in non-Hermitian quasiperiodic lattices}

\author{Ananya Ghatak}
\email{gananya@iesl.forth.gr}
\affiliation{Institute of Electronic Structure and Laser (IESL), Foundation for Research and Technology-Hellas (FORTH), P.O. Box 1527, 71110, Heraklion, Greece}

\author{Dimitrios H. Kaltsas}
\email{dkaltsas@physics.uoc.gr}
\affiliation{Institute of Theoretical and Computational Physics (ITCP), Department of Physics, University of Crete, 70013, Heraklion, Greece}

\author{Manas Kulkarni}
\email{manas.kulkarni@icts.res.in}
\affiliation{International Centre for Theoretical Sciences, Tata Institute of Fundamental Research,
Bangalore 560089, India}

\author{Konstantinos G. Makris}
\email{makris@physics.uoc.gr}
\affiliation{Institute of Electronic Structure and Laser (IESL), Foundation for Research and Technology-Hellas (FORTH), P.O. Box 1527, 71110, Heraklion, Greece}
\affiliation{Institute of Theoretical and Computational Physics (ITCP), Department of Physics, University of Crete, 70013, Heraklion, Greece}

\date{\today}

\begin{abstract}

Wave dynamics in disordered open media is an intriguing topic, and has lately attracted a lot of attention in non-Hermitian physics, especially in photonics. In fact, spatial distributions of gain and loss elements are physically possible in the context of integrated photonic waveguide arrays. In particular, in these type of lattices, counter-intuitive quantized jumps along the propagation direction appear in the strong disorder limit (where all eigenstates are localized) and they have also been recently experimentally observed. We systematically study the non-Hermitian quasiperiodic Aubry-André-Harper model with on-site gain and loss distribution (NHAAH), with an emphasis on the spectral sensitivity based on pseudospectra analysis. Moreover, diffraction dynamics and the quantized jumps, as well as, the effect of saturable nonlinearity, are investigated in detail. Our study reveals the intricate relation between the nonlinearity and non-Hermiticity.

\end{abstract}

\maketitle
\section{Introduction}
\label{sec:intro}

The interplay between disorder and wave dynamics is at the heart of the paradigmatic phenomenon of Anderson localization \cite{Anderson1,Anderson2}. In this framework the connection between the transport and localization of eigenstates due to disorder, has direct consequences for conductivity in metals. Beyond the area of condensed matter physics \cite{book1,book2}, that this key concept was initially introduced, Anderson localization has deeply influenced almost all areas of physics where disorder and waves meet. Few such examples include disordered photonics \cite{disphot1,disphot2,disphot3,disphot4}, localized phonons \cite{book3} and cold atom lattices \cite{book3}. 

The central cause of localization is the presence of randomness in a medium. In the general case of disorder, analytical treatment to a single realization of disorder is usually very difficult if not impossible, and thus alternative models have been introduced. The most representative case of such an analytically tractable model is the Aubry-André-Harper (AAH) model \cite{aah0,aah1,aah2}. This lattice provides an excellent platform representing a one-dimensional (1D) quasicrystal, which has attracted a lot of interest both theoretically and experimentally for the past several decades. The most striking feature of the AAH model is that, for an ideal incommensurate potential \cite{aah1,aah2}, the system undergoes a metal-insulator phase transition at a finite critical value of the quasiperiodic potential strength. Moreover, all localized eigenstates have the same localization length, which is a distinctive feature as compared to Anderson localization in disordered one-dimensional (1D) lattices. At the critical point, the energy spectrum is governed by the Harper equation, which describes the motion of a quantum particle on a two-dimensional crystal subjected to a magnetic flux and shows the characteristic Hofstadter butterfly energy spectrum \cite{aah3,aah4}.

On the other hand, openness due to either dissipation, material dispersion or coupling to an external environment is present in almost all systems. Thus the need for including dissipation in Anderson model was absolutely necessary. Apart from the random laser physics \cite{laser1,laser2,laser3} community - example where the strong nonlinearities undermines or hides the underlying physics,  the most relevant study towards this direction was the Hatano-Nelson lattice \cite{hatano1}. In this case, non-symmetric couplings were considered, and this asymmetry led to non-Hermiticity. This generalization of Anderson model attracted considerable attention \cite{hatano1,hatano2,hatano3,hatano4,hatano5}, and it was until recently that was experimentally observed in several, mainly optical, physical platforms. \cite{exp1,exp2,exp3,exp4,exp5}. 

Another alternative route to non-Hermiticity, that was recently introduced, was by combining actual materials that have gain and/or loss in order to create a composite system \cite{PT_review,PT1,PT2,PT3,PT4,PT5}. This idea that spanned non-Hermitian photonics, was triggered by the concept of $\mathcal{PT}$-symmetry \cite{Bender1,Bender2}, and became an active research area \cite{prb24,prb25,prb26,prb27,prb28,prb29,prb30,prb31,prb32,prb33,prb34,prb35,prb36,prb37,prb38,prb39,evenodd1,evenodd2,evenodd3}. Thus now by merging the two main constituents of Anderson localization, namely randomness and non-Hermiticity, a new twist in the fundamental Anderson localization problem was recently put forward \cite{CI1,CI2,CI3,jumps1,jumps2,jumps3}. Two representative effects that rely on engineered gain-loss spatial profiles in complex media and are directly related towards this new direction, are the constant-intensity waves \cite{CI1,CI2,CI3} and the beyond Anderson quantized jumps \cite{jumps1,jumps2,jumps3,ref3}. 

In this paper, moving away from random models,  we 
systematically examine an alternative platform for probing 
physics of disorder media, namely the quasiperiodic Aubry-
André-Harper model with on-site gain and loss distribution (NHAAH)\cite{lon1,lon2,lon3,quasi1,new3}. Such type of quasiperiodic models exhibit triple phase transition ($\mathcal{PT}$-symmetry breaking, topological-trivial, and localized-extended)\cite{quasi2}. We thoroughly explore the spectral properties and sensitivity based on pseudospectra analysis. In particular, we identify the existence of many pairs of exceptional points (EPs) for lattices with even number of sites (waveguide channels).  
Secondly, linear diffraction dynamics at the two different phases is examined and a systematic explanation of the quantized jumps based on the degree of non-orthogonality of the localized eigenstates, is provided. Furthermore, we consider the effect of saturable nonlinearity, which is typical in most laser systems, in the quantized jump dynamics.

The paper is organized as follows. In Sec.~\ref{sec:model}, we describe the details of the model considered and discuss its spectral properties. We also compute the phase rigidity \cite{rotter}, which is a quantitative measure for the non-orthogonality of the eigenfunctions of the non-Hermitian system, in order to characterize the appearance of EPs. In Sec.~\ref{sec:sens}, we 
study the sensitivity in the spectrum around the $\mathcal{PT}$-symmetric transition using pseudospectra analysis \cite{trefethen1,trefethen2,trefethen3,ps1,ps2,ps3}. In Sec.~\ref{sec:diff}, we discuss the linear diffraction dynamics and the appearance of quantized jumps \cite{new1, new2}. Sec.~\ref{sec:non} is dedicated to study the effects of nonlinearity and its interplay with non-Hermiticity. In Sec.~\ref{sec:conc} we provide the summary along with an outlook. Certain details are relegated to the appendices. 

\begin{figure}[h]
\hspace*{-0.5 cm}
\includegraphics[width=0.46\textwidth]{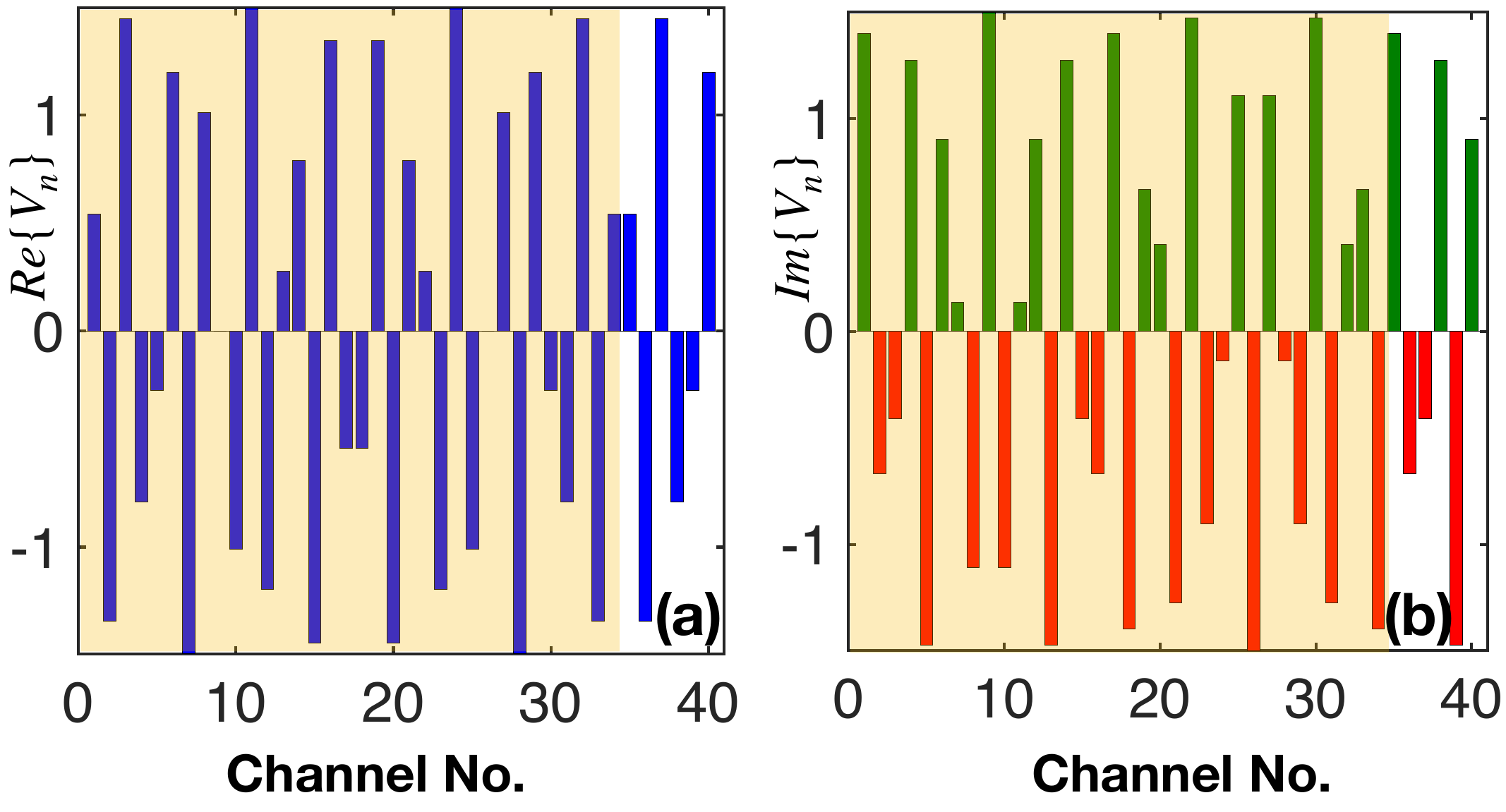}
\caption{(a) Real and (b) imaginary parts of the nearly incommensurate potential given in Eq.~\eqref{V}, as a function of the site index $n$ (channel number). In (b) the red-green colors correspond to gain and loss, respectively. The yellow shaded region in both plots represents the system size $N=34$ with $\alpha=21/34$, after which the potential repeats itself.}
 \label{potential}
\end{figure}

%%%%%%%%%%%%%%%%%%%%%%%%%%%%%%%%%%%
\section{Model and spectral properties of $\mathcal{PT}$-symmetric quasiperiodic potentials}
\label{sec:model}

Let us start by discussing the quasiperiodic non-Hermitian Aubry-André-Harper (NHAAH) model with on-site gain and loss and its eigenvalue spectrum. 
The system under consideration is a one-dimensional lattice, with incommensurate potential and periodic boundary conditions. The $N$-sites lattice is described by the following Hamiltonian 
\begin{gather}
H \equiv J\sum_{n=1}^{N-1} \Big( |{n+1} \rangle \langle{n}|+|{n} \rangle \langle{n+1}|\Big)+\sum_{n=1}^{N} V_n |{n} \rangle \langle{n}|\, ,
\label{Hpsi}
\end{gather}
where $n$ is the site index and $|{n} \rangle$ is the vector in local/position basis, $V_n$ is the on-site potential at each lattice site and $J$ is the coupling strength between neighbouring waveguides. The corresponding coupled mode equations of motion are:
\begin{gather}
i\frac{d\psi_n}{dz}+ J(\psi_{n-1}+\psi_{n+1})+V_n\psi_{n} =0\, . 
\label{H}
\end{gather}
where $\psi_n$ is the amplitude of the field's envelope at the $n^{th}$ site, under the paraxial approximation.
 The potential $V_n$ is of Aubry-Andre-Harper type that is $\mathcal{PT}$-symmetric \cite{lon1,lon2} and is given by
\begin {equation}
V_n= V_0\,e^{- 2 i  \pi \alpha n+i\pi \alpha (N+1)}\, ,
\label{V}
\end{equation}
where $V_0$ is a real constant that describes the strength of the non-Hermitian potential, $\alpha$ is a nearly incommensurate number, whereas $n$ is site index with $n=1,...,N$. We note that that the above potential is $\mathcal{PT}$-symmetric since it satisfies $Re\{V_{(N+1)/2+m}\}  = Re\{V_{(N+1)/2-m}\}$ and $Im\{V_{(N+1)/2+m}\}  = -Im\{V_{(N+1)/2-m}\}$ for odd $N$. In this case, $m=0,...(N-1)/2$. Similarly when $N$ is even the relation $Re\{V_{N/2-m}\}  = Re\{V_{N/2+(m+1)}\}$ and $Im\{V_{N/2-m}\}  = -Im\{V_{N/2+(m+1)}\}$ is satisfied  and in this case $m=0,1,2 \cdots N/2-1$.

To describe the quasiperiodic potential, we approximate $\alpha$ to the golden ratio, namely $\alpha=f_{i-1}/f_i$ where $f_i$ is the $i^{th}$ Fibonacci number.  The periodicity of the system requires the size of the system to be $N=f_i$, i.e., the system size needs to be chosen as one of the Fibonacci numbers. Here we consider the system size to be any of the numbers $N=f_i=21,34, 55, 89, 144, 233, 377, 610$. The quasiperiodicity is due to the incommensurate nature of $\alpha$, for the chosen system size. In other words, for lattice sizes higher than the above Fibonacci numbers, the system becomes periodic. To elaborate on this point, in Fig.~\ref{potential}, the real and imaginary parts of the complex potential are plotted for $N=34$, and we can clearly see the $\mathcal{PT}$-symmetric nature of the potential, as well as, its quasiperiodcity. After the yellow shaded area, the potential repeats itself in both real and imaginary parts.

\begin{figure}[h]
\hspace*{-0.5 cm}
\includegraphics[width=0.5\textwidth]{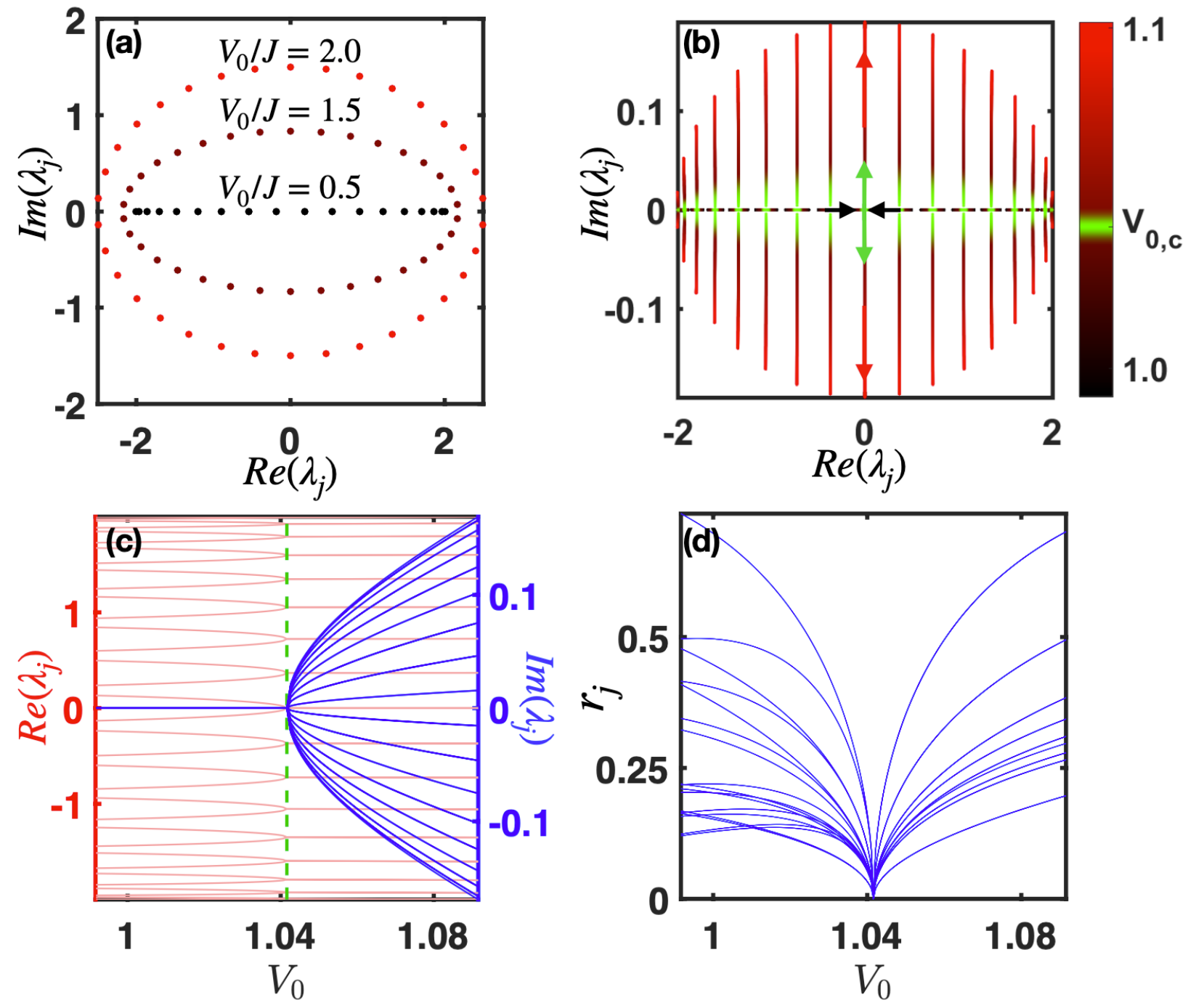}
\caption{Eigenspectra for even lattices. (a) Spectrum of the system in the complex plane for different values of  $V_0/J$, before and after $\mathcal{PT}$-phase transition for $N=34$ with $\alpha=21/34$. (b) Spectral flows (paths of complex eigenvalues in the complex plane) as we vary $V_0$ (see colorbar). Black, green and red arrows depict the trajectories of eigenvalues in three regimes of $V_0$, i.e., below, in the vicinity of, and above the EPs. (c) The real parts (red lines) and imaginary (blue lines) parts of eigenvalues as functions of $V_0$. The vertical green dashed line marks the $\mathcal{PT}$-phase transition point. (d) The phase rigidity is plotted versus $V_0$ with $J=1$.}
\label{even_spectrum}
\end{figure}

\begin{figure}[h]
	%\begin{center}
\hspace*{-0.5 cm}
\includegraphics[width=0.5\textwidth]{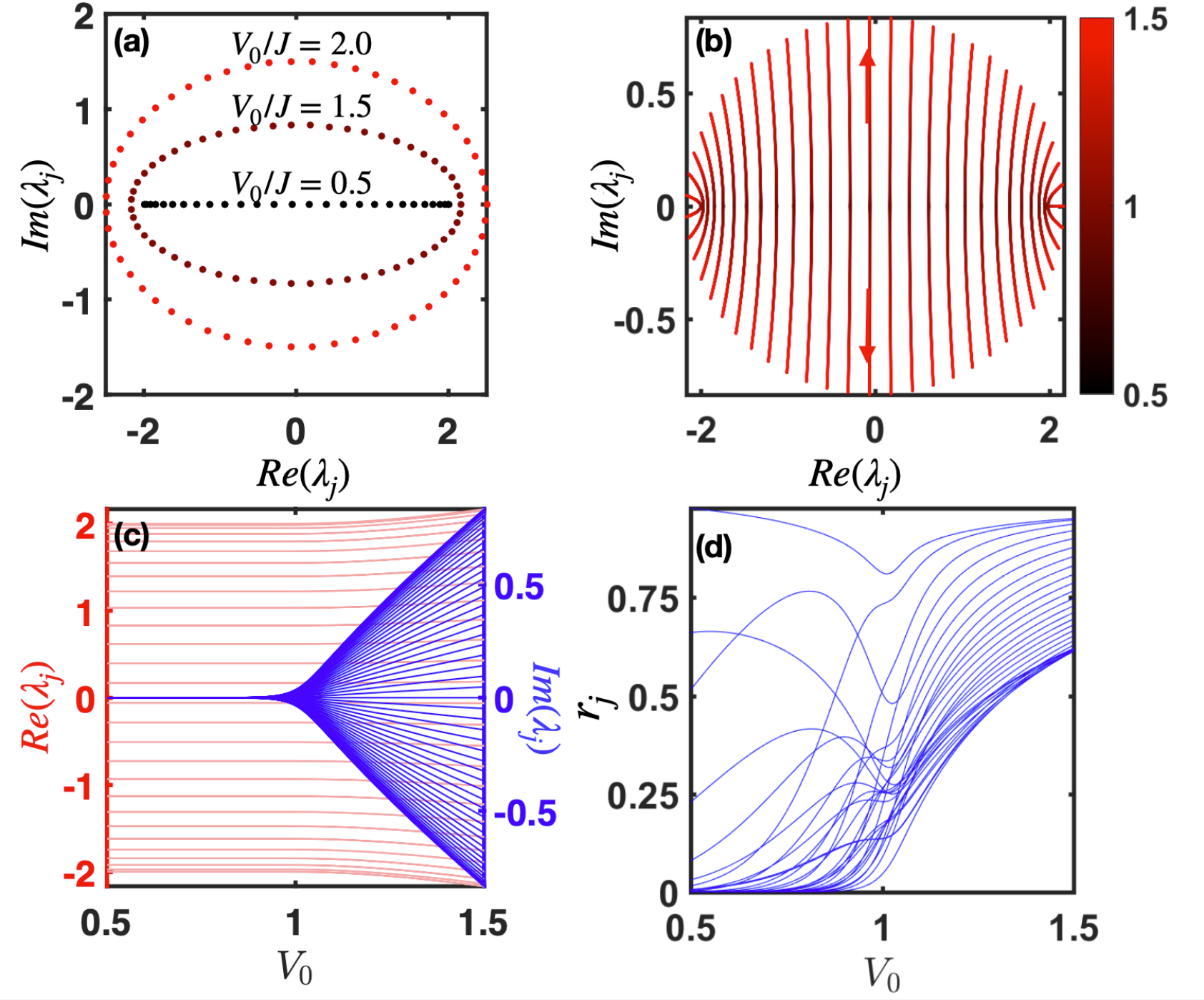}

\caption{Eigenspectra for odd lattices. (a) Spectrum of the system in the complex plane for different values of  $V_0/J$, for $N=55$ with $\alpha=34/55$. (b) Eigenvalue spectral flows in the complex plane, as we vary $V_0$ (see colorbar). Arrows depict the trajectories of eigenvalues. (c) The real parts (red lines) and imaginary (blue lines) of eigenvalues are plotted versus $V_0$. (d) The phase rigidity of the system has been plotted versus $V_0$ with $J=1$.}
	%\end{center}
 \label{odd_spectrum}
\end{figure}

The incommensurate nature of $\alpha$, has interesting consequences like an abrupt phase transition \cite{lon1,lon2} even in the one  dimension, in contrast to the random Anderson model. In order to examine such a transition, we first have to study the eigenspectrum of the lattice. The eigensolutions of the (right) eigenvalue problem associated with this Hamiltonian in Eqs.~(\ref{H},\ref{V}) by assuming $\psi_{n}(z)=u^R_{j,n} \,e^{i\lambda_jz}$ 
is 
\begin{equation}
H\vert u_j^R \rangle=\lambda_j \vert u_j^R \rangle\, ,
\label{eq:hright}
\end{equation}
where $\lambda_j $ is the complex eigenvalue of the system. The right eigenvalue 
problem in Eq.~\eqref{eq:hright} with right eigenvectors ($\vert u_j^R \rangle$), together with the corresponding left eigenvalue problem with the left eigenvectors ($\vert u_j^L \rangle$), that satisfy the biorthogonality relation $\langle u_i^L\vert 
u_j^R\rangle =\delta_{ij}$, provide a basis for expanding any initial excitation. Based on these definitions, the eigenspectra for even ($N=34$) and odd ($N=55$) lattices are shown in Fig.~\ref{even_spectrum}, and Fig.~\ref{odd_spectrum}, respectively.

In particular, regarding the $N$-even lattice, for $V_0/J<1$, the system has real eigenvalues as shown in Fig.~\ref{even_spectrum}(a), as a consequence of the unbroken $\mathcal{PT}$-symmetry. On the other hand, when  $V_0/J>1$, all eigenvalues form complex conjugate pairs as the result of broken $\mathcal{PT}$-symmetry. Apart from this $\mathcal{PT}$-phase transition, the system exhibits different behavior for different values of $V_0/J$-ratio. For low values the eigenstates are delocalized whereas for high values are localized. At the limit of large lattices with $N$-even ($N \to \infty$) there is a special point at $V_{0,C}/J \equiv V_0/J=1$, at which the system exhibits a simultaneous phase transition from unbroken $\mathcal{PT}$-symmetry to broken and from extended to localized eigenstates (see Appendix ~\ref{app:pt_tp}). To elucidate the subtle point of the EPs formation, we plot the spectral flows of the eigenvalues in the complex plane for an even sized system with respect to $V_0$ in Fig.~\ref{even_spectrum}(b). The arrows show how at the critical $V_{0,C}$ the eigenvalues coalesce and form many pairs of second-order EPs during the phase transition. Note that in this case $V_{0,C}$ is not equal to one since the system size chosen is still far from the large-$N$ limit (see Appendix ~\ref{app:pt_tp}).  In Fig.~\ref{even_spectrum}(c) we plot the real (red lines) and imaginary (blue lines) parts of the eigenvalues $\lambda_j$ with respect to $V_0/J$. As we can see, the system exhibits an abrupt phase transition from unbroken $\mathcal{PT}$-symmetry to a broken one, that is a typical feature of these type of Hamiltonians. 

On the contrary, for $N$-odd the picture is different. The $\mathcal{PT}$-symmetry is always broken \cite{evenodd1,evenodd2,evenodd3} and lattice's spectrum is partially real and partially complex, as is shown in Fig.~\ref{odd_spectrum} (a,b,c). The eigenvalues form complex conjugate pairs with slowly growing imaginary parts as we increase $V_0/J$ (as indicated by the arrows). The spectral flow in Fig.~\ref{odd_spectrum}(b) and bifurcation curves in Fig.~\ref{odd_spectrum}(c) show that the real parts of the eigenvalues do not approach each other, since they do not form EPs. The imaginary parts of the eigenvalues are close to each other but do not coalesce abruptly as in the case of even system's size. Regarding the localization-delocalization transition, that is evident at the limit of large lattice sizes. Here for the particular example of $N=55$ the eigenstates have similar behavior as in the even case, namely as the $V_0/J$ is increasing they become more localized.

To further corroborate the formation of multiple pairs of EPs, we calculate the phase rigidity which is a quantitative measure for the non-orthogonality of the associated eigenfunctions. Because the Hamiltonian is symmetric ($H = H^{T}$), the left and right eigenstates, respectively corresponding to the $\lambda_j$ and $\lambda^{*}_j$, are complex conjugates of 
one another, i.e., $\vert u_j^R\rangle=\vert u_j^L\rangle^*$. 
The definition of phase-rigidity \cite{rotter,moiseyev,phr3} is given as,
\begin {equation}
\label{eq:pr}
r_j = \frac{|\langle u_j^L | u_j^R \rangle|}{\langle u_j^R | u_j^R \rangle}= \frac{1}{\langle u_j^R | u_j^R \rangle}.
\end {equation}
In Fig.~\ref{even_spectrum}(d) and Fig.~\ref{odd_spectrum}(d) we plot the phase rigidity of all the eigenstates as a function of the potential strength for even and odd $N$, respectively. For even cases, when the system exhibits pairs of EPs, the phase rigidity of the associated eigenstates becomes exactly zero, as it can be seen in Fig.~\ref{even_spectrum}(d). For odd case shown in Fig.~\ref{odd_spectrum}(d), the phase rigidity varies between zero and one. But there is no sudden drop of phase rigidity to zero, since we do not have formation of EPs.

%%%%%%%%%%%%%%%%%%%%%%%%%%%%%%%%%%%%%%%%%%%%%%%%
\section{Spectral Sensitivity based on pseudospectra}
\label{sec:sens}

One of the intriguing characteristics of $\mathcal{PT}$-symmetry phase transitions is the formation of EPs, which are highly sensitive to external perturbations. As discussed in Sec.~\ref{sec:model}, the even and odd lattices exhibit different kinds of behaviour as we vary $V_0$. Here, we computationally examine the sensitivity of the even and odd lattices by using the general framework of \textit{pseudospectra} \cite{trefethen1}. The concept of pseudospectrum is known and applied in various areas of mathematics \cite{trefethen2,trefethen3}, but apart from fluid mechanics, is largely unknown in physics \cite{ps1,ps2,ps3}. By construction, such generalized version of the eigespectrum is ideal for non-Hermitian problems, where the pseudospectrum is, in general, very different from the corresponding eigenspectrum \cite{trefethen1}. Additionally, the pseudospectra of an associated matrix, contain valuable information regarding both the sensitivity and the dynamics of the corresponding problem. In this section, we compute the related pseudospectra by using three different, but equivalent methods \cite{trefethen1,trefethen2,trefethen3}.  

\textit{Method 1-Random matrix ensemble:} The first method relies on adding random perturbations $E$ with strength $\epsilon$ to the Hamiltonian $H$. For a large number of perturbations, this method provides a direct picture of the sensitivity of the Hamiltonian. In other words, the $\epsilon$-pseudospectrum $\sigma_\epsilon (H)$ of the Hamiltonian is related to the eigenvalue spectrum of the perturbed matrix $H + E$ with $||E|| \le \epsilon$. The mathematical definition is
\begin{align}
   \sigma_\epsilon (H)\equiv \bigcup_{i=1,||E_i|| < \epsilon}^{M}\{\sigma (H+E_i)\ :\ ||E|| \le \epsilon \}\, ,
   \label{eq_pseudo}
\end{align}
where $M$ is the number of different realizations of the perturbations, $\sigma(H)$ denotes the eigenvalue spectrum of the Hamiltonian and $||...||$ is the matrix norm which is defined by $||A||=\underset{x\neq 0}{\sup}\frac{||Ax||}{||x||}$.

\textit{Method 2-Singular values:} A more precise way to calculate the $\epsilon$-pseudospectrum is by relying on singular values. The smallest singular value of the matrix $\mathbf{z}I-H$, denoted as $s_{\mbox{min}}(\mathbf{z}I - H)$, measures how close the complex number $\mathbf{z}\in C$ is to an eigenvalue. Thus, we define the $\epsilon$-pseudospectrum as:
\begin{align}
   \sigma_\epsilon (H)\equiv \{ \mathbf{z}\in C: s_{\mbox{min}} (\mathbf{z}I - H)\le \epsilon \}\, .
   \label{psmethod2}
\end{align}
Given a particular value of the perturbation strength $\epsilon$, this method provides accurate limits of the possible value of the eigenvalues of the perturbed matrix.

\textit{Method 3-Resolvent:} An alternative for calculating the $\epsilon$-pseudospectrum is the norm of the matrix resolvent, namely the $||(\mathbf{z}I-H)^{-1}||$. When $\mathbf{z}$ coincides with an eigenvalue, the matrix $\mathbf{z}I-H$ is not invertible, whereas in the vicinity of the eigenvalues, the norm $||(\mathbf{z}I-H)^{-1}||$ becomes large. The mathematical definition is
\begin{align}
   \sigma_\epsilon (H)\equiv \{ \mathbf{z}\in C: ||(\mathbf{z}I-H)^{-1}||\ge \epsilon^{-1} \}\, .
   \label{psmethod3}
\end{align}
All three methods are equivalent and we use them all in the pseudospectra calculations that follows. Furthermore, the size of the pseudospectrum corresponding to an eigenvalue or a pair of eigenvalues forming an EP2, can be quantitatively described by the local pseudospectral radius $\rho_\epsilon$ \cite{trefethen1,ps3}, which is defined as 
\begin{equation}
\label{eq:cloud}
\rho_\epsilon\equiv \mbox{max}_{\mathbf{z} \in B}Im(\mathbf{z}) \, , 
\end{equation}
where $B$ is a subset of $\sigma_\epsilon(H)$. For different $\epsilon$-pseudospectra we can compute the corresponding $\rho_\epsilon$ and find its functional dependence with $\epsilon$. A typical characteristic feature of an EP2, based on non-Hermitian perturbation theory, is a square root response $\rho_\epsilon\propto\epsilon ^{1/2}$, whereas eigenvalues of normal matrices exhibit a linear response $\rho_\epsilon\propto\epsilon$ \cite{ps3}.

\begin{figure}[h]
	%\begin{center}
\hspace*{-0.5 cm} 
\includegraphics[width=0.50\textwidth]{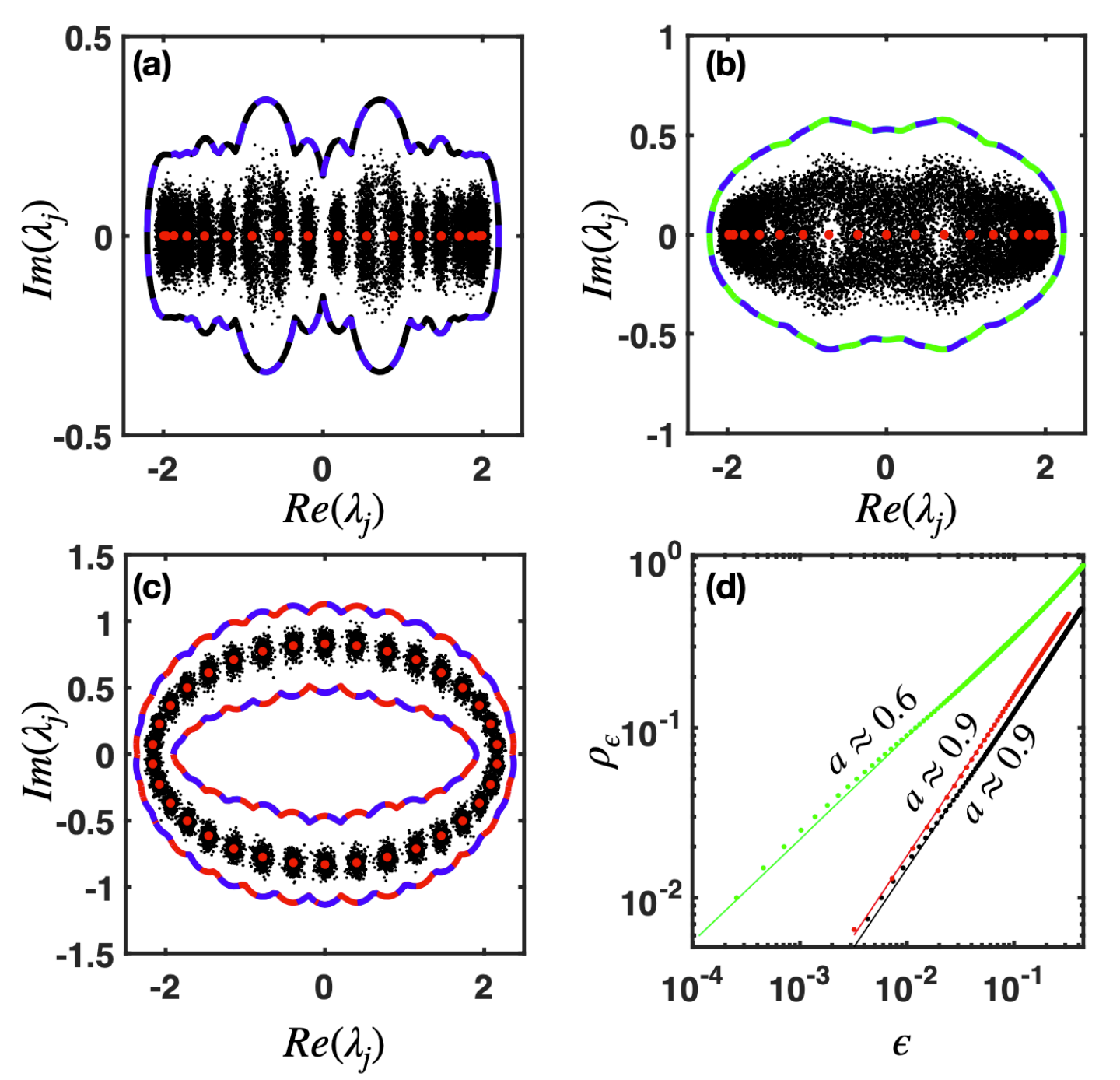}

\caption{(a),(b),(c) Pseudospectra $\sigma_{0.2}(H)$ for an even lattice ($N=34$), before ($V_0/J=0.5$), at ($V_0/J=V_{0,C}/J\approx 1.04$) and after ($V_0/J=1.5$) the phase transition point, with all three methods (black dots, black-blue, blue-green, blue-red lines), respectively. The red dots show the eigenspectrum. (d) $\rho_\epsilon$ given in Eq.~\eqref{eq:cloud} is plotted for the case of (a),(b) and (c) as a function of perturbation strength. Dots are data and lines are the linear fits. The slopes are 0.9 (black), 0.6 (green) and 0.9 (red) respectively.} 
	%\end{center}
 \label{even_ps}
\end{figure}

Our results regarding the pseudospectra analysis of the even-sized systems are shown in Fig.~\ref{even_ps}. In Fig.~\ref{even_ps}(a,b,c) we calculate the pseudospectrum before, at and after the phase transition, respectively. The black dots are the eigenvalues of the perturbed matrix according to the mathematical definition in Eq.~(\ref{eq_pseudo}), which we refer to as method-1. The black, green and red lines are the results of method-2, respectively i.e., the singular values method given in Eq.~\eqref{psmethod2}. The blue lines are the results of the resolvent approach [see Eq.~\eqref{psmethod3}] which we refer to as method-3. These lines accurately limit the eigenvalues for a given perturbation strength $\epsilon$. All three methods are in excellent agreement with each other. The corresponding eigenspectra are also shown (red dots) for comparison. Moreover, in Fig.~\ref{even_ps}(d) we calculate the local pseudospectral radius of the eigenstate closest to $Re(\lambda_j)=0$ (away from the two peaks) before, during and after the phase transition. By fitting we can express the sensitivity of the even lattice systems as $\lim_{\epsilon \to 0} \rho_\epsilon\propto \epsilon ^a$ and numerically determine the exponent $a$. For $V_0<V_{0,C}\approx J$ and the even-sized lattices we find $a\approx0.9$ which is indicative of typical sensitivity (almost linear). For $V_0=V_{0,C}$ we find a square root behavior ($a\approx0.6$) which demonstrates the ultra-sensitivity of our lattice close the phase transition point. For $V_0>V_{0,C}$, we find a nearly linear response to external perturbations ($a\approx 0.9$).

\begin{figure}[h]
	%\begin{center}
\hspace*{-0.5 cm} 
\includegraphics[width=0.50\textwidth]{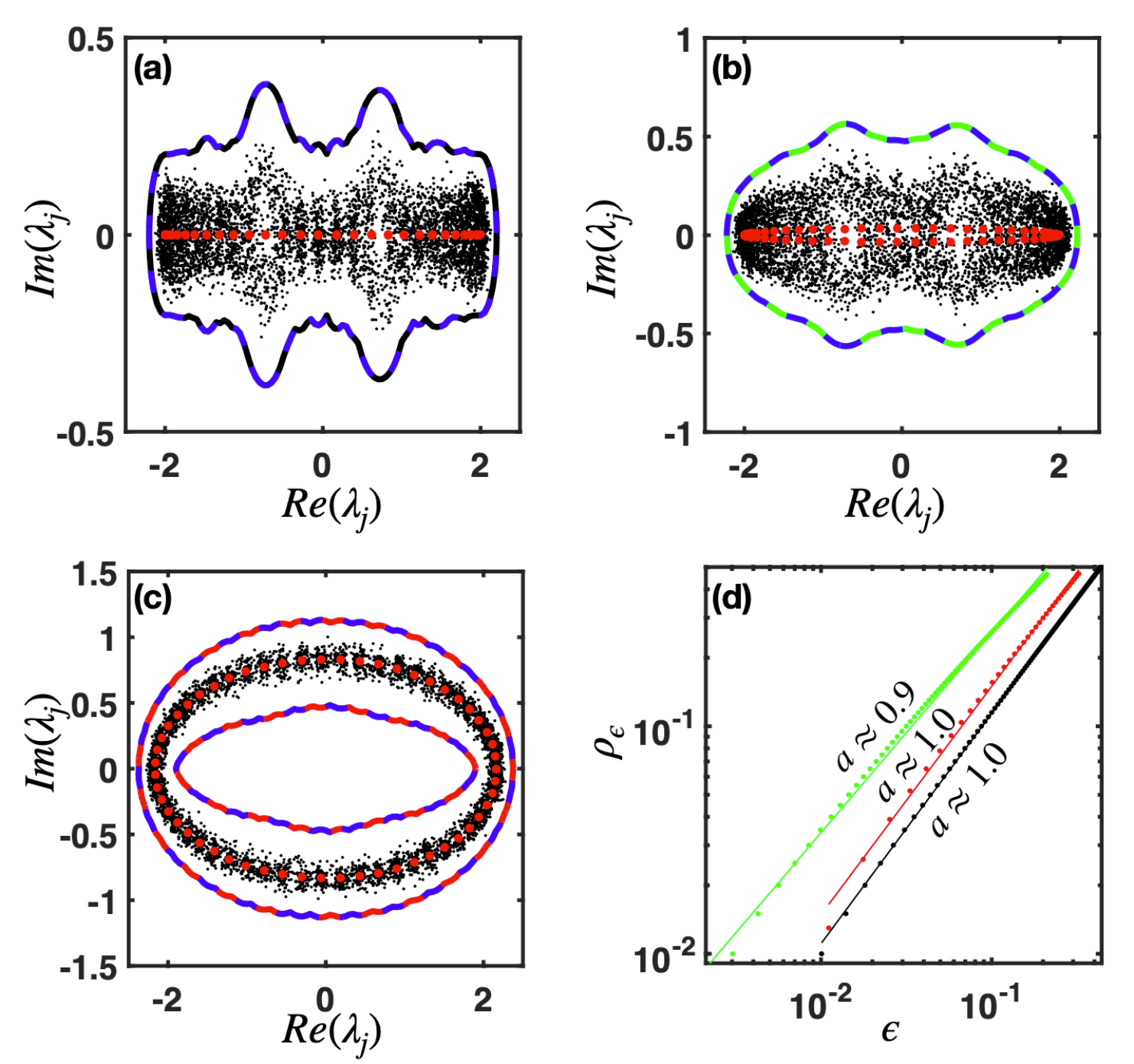}

\caption{(a),(b),(c) Pseudospectra $\sigma_{0.2}( H)$ for an odd lattice ($N=55$) for ($V_0/J=0.5$), ($V_0/J=1$) and ($V_0/J=1.5$), with all three methods (black dots, black-blue, blue-green, blue-red lines) respectively. The red dots show the eigenspectrum. (d) $\rho_\epsilon$ given in Eq.~\eqref{eq:cloud} are plotted for the case of (a),(b) and (c) as a function of perturbation strength. Dots are data and lines are the linear fits. The slopes are 1.0 (black), 0.9 (green) and 1.0 (red) respectively.}
	%\end{center}
 \label{odd_ps}
\end{figure}

A similar analysis is performed for the odd-sized systems in Fig.~\ref{odd_ps}, where Fig.~\ref{odd_ps}(a,b,c) contain the pseudospectra based on the three definitions before, for three different values of $V_0/J$, respectively. For $V_0/J<1$, the sensitivity of the eigenvalues depends on the position of their real part on the real axis $Re(\lambda_j)$. In the limit, $N\rightarrow\infty$, both pseudospectra of even and odd-sized lattices exhibit two peaks around $Re(\lambda_j)\approx\pm 1$, which have the same sensitivity (square root behavior) even without the presence of EPs, whereas the sensitivity away from these peaks is linear. Thus, even and odd lattices exhibit the same sensitivity before and after $V_0/J=1$. However, this fact is not true for $V_0/J=1$. More specifically, in Fig.~\ref{odd_ps}(d) the local pseudospectral radius of the eigenstate closest to the origin is plotted, before, at and after the $V_0/J=1$. For all three cases the system exhibit linear response to external perturbations ($a\approx1$), which is consistent with the fact that the odd lattice does not have EPs.

%%%%%%%%%%%%%%%%%%%%%%%%%%%%%%%%%%%%%%%%%%%%%%%%%%%%%%%
\section{Diffraction Dynamics}
\label{sec:diff}

In this section, we provide a comprehensive analysis of the wave propagation in quasiperiodic non-Hermitian lattices. In general and at the large lattice size limit, for $V_0/J<1$ all the eigenstates are extended, hence, we anticipate wavepacket spreading behavior in the diffraction pattern. Conversely, for $V_0/J>1$, where the eigenstates are localized, we expect to observe the effects of eigenstate localization in the diffraction. These behaviors correspond to metallic and insulating phases respectively. However, recent works have shown that the interplay between non-Hermitian optics and Anderson localization may result in abrupt quantized ``jumps"~\cite{jumps1,jumps2,jumps3}.

\begin{figure}[h]
\hspace*{-0.4 cm} \includegraphics[scale=.32]{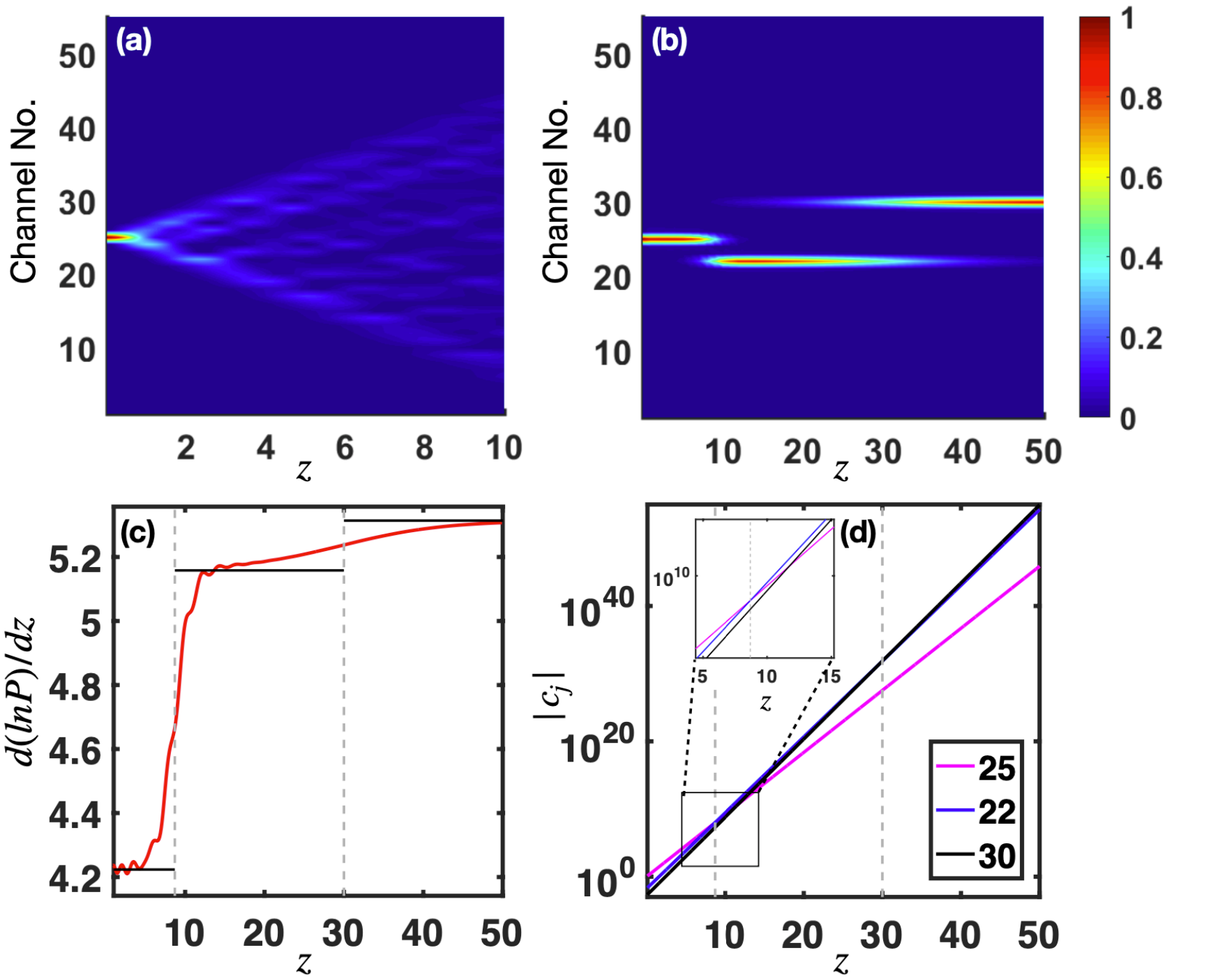}
\caption{(a),(b) Dynamical evolution under single channel excitation $\psi_n(z=0)=\delta_{n,25}$ for ($V_0/J=0.5$) and ($V_0/J=3$), respectively. (c) Derivative of the power logarithm (red line) over $z$ for the dynamics shown in (b). The dashed gray lines correspond to the propagation $z$ where the jumps happen and the black lines are the analytical prediction[see Eq.~\eqref{logP}]. (d) Magnitude of $|c_j(z)|$ for the eigenstates that participate in the jumps of (b). The inset is a zoom near the first jump. The indices indicate the localization sites of the eigenstates. The parameters used were $N=55$ and $\alpha=34/55$.}
\label{dyn_l}
\end{figure}

To further analyze the dynamics, we study the evolution of the quasiperiodic system described by Eq.~(\ref{H}). Our analysis begins with a key physical quantity, the optical power $P(z)$ at propagation distance $z$, which is
defined as
\begin {equation}
P(z) =\sum_{n=1}^N|\psi_n(z)|^2\, .
\label{pz}
\end {equation}
For $V_0/J>1$, the field amplitude $|\psi_n (z)|$ diverges exponentially as $z \rightarrow \infty$ due to the high non-zero values of the imaginary parts of the eigenvalues. Therefore, we introduce the normalized field amplitude $\phi_n(z) = \frac{\psi_n(z)}{\sqrt{\it{P}(z)}}$.
Although somewhat different from the field amplitude, the normalized field amplitude retains the same information and accurately describes the relevant wavepacket's position. The normalized field intensities $|\phi_n(z)|^2$ for $V_0/J=0.5$, and $V_0/J=3$ are plotted in Fig.~\ref{dyn_l}(a,b) as a function of the propagation distance $z$ for single channel excitation. Before the $V_0/J=1$ point [Fig.~\ref{dyn_l}(a)], the wave propagates by coupling to the adjacent waveguide channels. After the $V_0/J=1$ point [Fig.~\ref{dyn_l}(b)], the eigenstates are localized and the system is in the insulating phase. Additionally, it exhibits two jumps between the sites $25$,$22$ and $30$ at $z=9$ and $z=30$ respectively.

The physical mechanism that enables the jumps is  the competition between different localized eigenmodes with complex eigenvalues. Specifically, eigenmodes associated with eigenvalues that have a large imaginary part can dominate the dynamics due to their stronger influence. As these dominant eigenmodes overpower the others, they facilitate the transport phenomenon observed in the system, i.e. the quantized jumps.

To further investigate the jumpy behavior, we adopt the methodology of previous related work \cite{jumps3}. We begin by expanding the field amplitude $\psi_n$ in a linear combination of the right eigenstates as
\begin {equation}
\psi_n(z) =\sum_{j=1}^N c_j(z) u_{jn}^R\, ,
\label{psi}
\end {equation}
where $c_j(z)=\langle u_j^L|\psi (z)\rangle$ are the projection coefficients and $u_{jn}^R$ is the $n^{th}$ element of the eigenvector $|u_j^R\rangle$. Using Eq.~(\ref{psi}) we can derive a modal form for $P(z)$, based on $\mathcal{O}_{i,j}=\langle u_{j}^R|u_{i}^R\rangle$, the overlap between two right eigenstates. For non-Hermitian Hamiltonians, the right eigenstates are not generally orthogonal to each other. However, localized eigenstates tend to have minimal overlap when they are sufficiently separated. When the eigenstates are highly localized they are almost orthogonal to the other eigenstates and this can be enough to semi-analytically calculate the jump positions. This assumption is referred to as ``almost orthogonal" approximation ~\cite{jumps3} and we have considered a relevant discussion to our problem in Appendix \ref{sec:overlap}. By using the almost orthogonal approximation, we can write
\begin {equation}
\mathcal{O}_{i,j} \approx \delta_{i,j}\mathcal{O}_{j,j}\, ,
\label{Gamma}
\end {equation}
Additionally, the localization of the dynamics implies that for specific intervals along the propagation axis $z$, one eigenmode $|u_m^R\rangle$ dominates the dynamics, i.e. its projection coefficient is much higher than the rest. In other words, $|c_m|\gg |c_j|$. During the $z$ interval where the eigenstate with eigenvalue $\lambda_m$ is dominant, that single mode determines the derivative of the logarithmic power. Therefore, we have:
\begin {eqnarray}
\frac{d(ln P(z))}{dz}\approx -2Im(\lambda_m\,) .
\label{logP}
\end {eqnarray}
This measurable quantity of the logarithmic power can accurately predict the location, duration and number of the jumps as is evident from Fig.~\ref{dyn_l}(c). In particular, we plot the derivative of the logarithmic power with respect to the propagation distance $z$, for the jumpy dynamics of Fig.~\ref{dyn_l}(b). The black lines are the theoretical predictions $-2Im(\lambda_m)$ where $\lambda_m$ corresponds to the eigenstates involved in the jumps. Specifically for our example, these are the states localized at the sites $25$,$22$ and $30$ during the first $0<z<9$ (initially), second $9<z<30$ (after the first jump) and third $z$ interval, $z>30$ (after the second jump). The analytical and numerical predictions are in excellent agreement.

Next we provide an complementary way to study the jumps using the projection coefficients of the eigenmodes $|c_j(z)|$ to the initial condition. The coefficients obey the relation $|c_j(z)|=|c_j(0)|e^{-Im(\lambda_j)z}$, which implies that their logarithm are straight lines i.e. $log(|c_j(z)|)\propto -Im(\lambda_j) z$. When the lattice is excited, some eigenmode coefficients may initially have small values. However, as they propagate, these modes may surpass those with higher initial value due to their larger slope [$-Im(\lambda_j)$]. The eigenmode with the largest coefficient, is referred to as the dominant mode. Subsequently, the eigenmode with the steepest slope becomes the new dominant mode due to its larger growth. In other words, the jump happens at the position $z$ where the line corresponding to the dominant eigenstate intersects another line. This is a precise way to define where the jump occurs, without relying on the dynamics of Fig.~\ref{dyn_l}(b). Our results are shown in Fig.~\ref{dyn_l}(d), where we plot the magnitude of the projection coefficients $|c_j(z)|$ of the three modes that participate in the jumps as a function of the propagation distance $z$. The inset highlights the first jump at the first gray dashed line at $z=9$. There, the mode localized at site 25 (purple line) has its projection coefficient surpassed by the eigenmode localized at site 22 (blue line). Then, at $z=30$, marked by the second gray dashed line, the second jump is taking place and the state becomes localized at site 30 (black line). Additionally, when the difference in slopes between the lines is very small, the jump is less abrupt. This is because the coefficients are ``almost equal" for a large interval $z$. For example, the pink and blue line have a larger difference in slopes than the blue and black line. As a result, we expect the first jump to be more abrupt than the second, a feature that can also be seen in Fig.~\ref{dyn_l}(b). By direct comparison of Fig.~\ref{dyn_l}(c) and Fig.~\ref{dyn_l}(d), we can conclude that the two different approaches, that of optical power and that of projection coefficients are in very good agreement and provide an explanation for the effect of quantized jumps. Also similar conclusions can be drawn for the case of quasiperiodic lattices with even number of waveguide channels.

%%%%%%%%%%%%%%%%%%%%%%%%%%%%%%%%%%%%%%%%%%%%%%%%%%%%%%%%%%%%
\section{Effect of saturable nonlinearity}
\label{sec:non}
\begin{figure}[h]
\includegraphics[scale=.34]{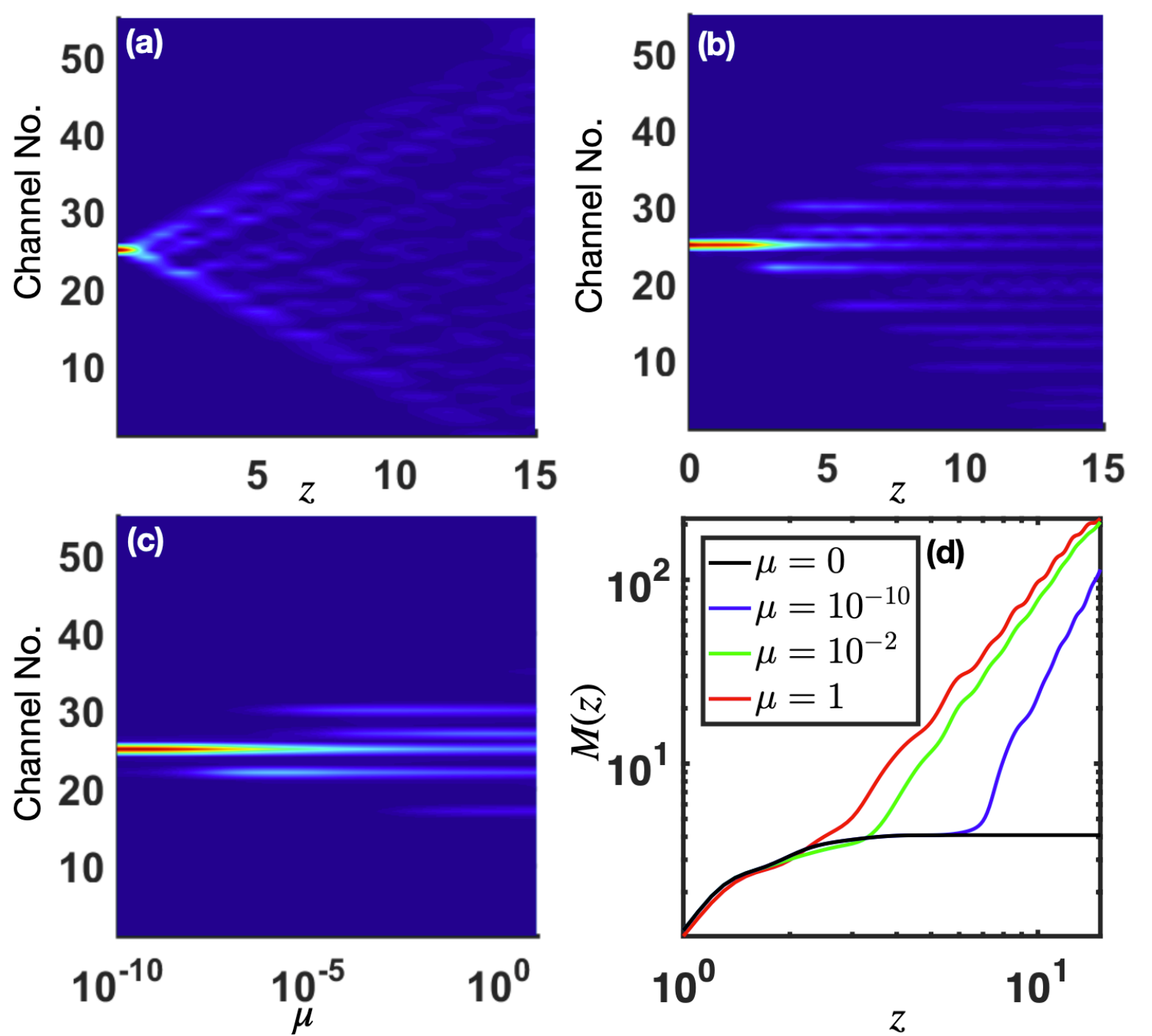}

\caption{Effect of nonlinear saturable gain. (a),(b) Evolution of the normalized field for single channel excitation under the influence of nonlinearity ($\mu=10^{-2}$), in the metallic ($V_0/J=0.5$)  and insulating phase ($V_0/J=3$), respectively. (c) The normalized intensity $|\phi_n(z=5)|$ for different values of the nonlinearity parameter, $\mu$ in the insulating regime $V_0/J=3$. (d) Evolution of variance for different values of $\mu$. The parameters used are $N=55$, $\alpha=34/55$ and the initial condition is always $\psi_n(z=0)=\delta_{n,25}$.}
	%\end{center}
  \label{dyn_nl}
\end{figure}

In the previous sections all discussions were devoted to the linear non-Hermitian quasiperiodic system. Here we are going to consider the effect optical nonlinearity, that exist in these types of photonic lattices. In particular, we  assume a saturable type of nonlinearity, that is typical in gain-laser materials. More specifically, the gain is nonlinear whereas the loss linear, thus the potential can be written now as follows:
\begin{equation}
V_n\rightarrow \tilde{V}_{n}(|\psi_n|)\equiv 
\begin{cases}
Re(V_n)+i\frac{ Im(V_n)}{1+\mu|\psi_{n}|^2}, & \text{if}\ Im(V_n)< 0 \\
Re(V_n)+iIm(V_n), & \text{if}\ Im(V_n) \ge 0\, . \\
\end{cases}
\end{equation}
Then, the dynamical evolution of the field is governed by
\begin {equation}
i\frac{d\psi_n}{dz}+J(\psi_{n-1}+\psi_{n+1})+\tilde{V}_n(|\psi_n|)\psi_{n}=0\, .
\label{Hnl}
\end {equation}
Notice that in the limit $\mu\rightarrow 0$, the saturable gain is the same as in Eq.~(\ref{V})($\tilde {V_n}(|\psi_n|)\rightarrow V_n$). In that limit, the dynamics are similar to the quasiperiodic system. For the nonlinear case where $\mu>0$, we plot the evolution of the normalized field intensity $|\phi_n(z)|^2$ in Fig.~\ref{dyn_nl}(a,b) below and above the point $V_0/J=1$. The initial power is always kept constant $P(0)=1$. As we can see for $V_0/J>1$, the nonlinearity prevents jumps, despite the localized eigenstates, in sharp contrast to the linear case. To see this explicitly, in Fig.~\ref{dyn_nl}(c) we also plot the state of the normalized field intensity $|\phi_n(z=5)|^2$ with respect to the nonlinear strength $\mu$ for a fixed propagation distance $z=5$.  This clearly shows the insulator to metallic behavior as the nonlinearity, $\mu$, increases. In order to quantify the spreading of the wavepacket along propagation we examine the variance $M(z)$, which  is defined as follows:
\begin {equation}
M(z) \equiv \sum_{n=1}^N (n - n_0 )^2 |\phi_n(z)|^2.
\label{mz}
\end {equation}
In  Fig.~\ref{dyn_nl}(d) we plot the variance with respect to propagation distance $z$ for different values of nonlinear strengths $\mu$. For $\mu=0$, all the states are localized at or near their initial position for a single channel excitation.  Therefore, $M(z)$ saturates for large values of $z$ ($M(z) \propto z^0$). As the nonlinearity increases the system enters the metallic phase and $M(z)$ increases.

%%%%%%%%%%%%%%%%%%%%%%%%%%%%%%%%%%%%%%%%%%%%%%%%%%%%%
\section{Conclusions} 
\label{sec:conc}

In summary, we have examined $\mathcal{PT}$-symmetric non-Hermitian quasiperiodic NHAAH-model, for even and odd number of sites. In particular, our study was focused on three aspects, namely (a) the sensitivity of the lattice based on pseudospectra, (b) quantized jumpy propagation around distant sites due to eigenstate localization \cite{jumps2,jumps3} and (c) the effect of saturable gain nonlinearity on the wavepacket's evolution. More specifically, pseudospectra revealed the enhanced sensitivity of the lattice on external perturbations, around EPs when they were present (even number of sites), but also when they were absent (odd number of sites), due to the underlying eigenstate non-orthogonality. Regarding the occurrence of quantized jumps, they were evident in both cases of even, odd lattices for high values of $V_0/J$, when localization was strong. The saturable gain nonlinearity was also not favouring jumps, but rather had a detrimental effect. At this point we comment on the experimental relevance of our non-Hermitian quasiperiodic lattices. In fact there are already related experiments regarding disordered systems in optical fiber loop networks \cite{jumps1}, which can be tailored to realize quasiperiodic variations of the optical potential. Furthermore, the discussed results can be physically implemented by using alternative experimental platforms including waveguide arrays~\cite{ref1,exp5} and coupled microring resonators~\cite{exp3}.

Our work revealed a possible deep connection between pseudospectra and appearance of quantized jumps in diffraction dynamics of quasiperiodic non-Hermitian systems. Further rigorous understanding of this intriguing relationship and its validity in a wider class of models, i.e., possible universality, is currently under investigation. 
It will also be interesting to explore spectral sensitivity and quantum dynamics of localized initial conditions in interacting non-Hermitian \cite{iq1,iq2,iq3} quasiperiodic and disordered quantum systems. In conclusion, we believe that our study may be highly relevant to wave transport in complex media that are non-Hermitian and random, for any model of randomness either disordered or quasiperiodic.

%%%%%%%%%%%%%%%%%%%%%%%%%%%%%%%%%%%%%%%%%%%%%%%%%%%%

\section{Acknowledgements} 

This project was funded by the European Research Council (ERC-Consolidator) under grant agreement No. 101045135 (Beyond Anderson). D. K. acknowledges support by the Hellenic Foundation for Research and Innovation (HFRI) under the 4th Call for HFRI PhD Fellowships (Fellowship Number: 10638.). M. K. thanks the hospitality of the Department of Physics, University of Crete (UOC) and the Institute of Electronic Structure and Laser (IESL) - FORTH, at Heraklion, Greece.  M. K. thanks the VAJRA faculty scheme (No.~VJR/2019/000079) from the Science and Engineering Research Board (SERB), Department of Science and Technology, Government of India. M. K. acknowledges support of the Department of Atomic Energy, Government of India, under project no. RTI4001.

\appendix
\renewcommand\thefigure{\thesection.\arabic{figure}}

\section{Scaling up the system}
\label{sec:scaling}

In this appendix, we will discuss various quantities as we scale up the system size. 
More specifically, we discuss the change in the value of the $\mathcal{PT}$-phase transition point $V_{0,C}$ in Appendix \ref{app:pt_tp}, as we scale up $N$ but keeping it even. In Appendix~\ref{sec:app_pseudo} we study the sensitivity via pseudospectra for larger system sizes with an even number of channels.

\setcounter{figure}{0}    
\begin{figure}%[h]
	%\begin{center}
\hspace*{-0.5 cm} \includegraphics[scale=.32]{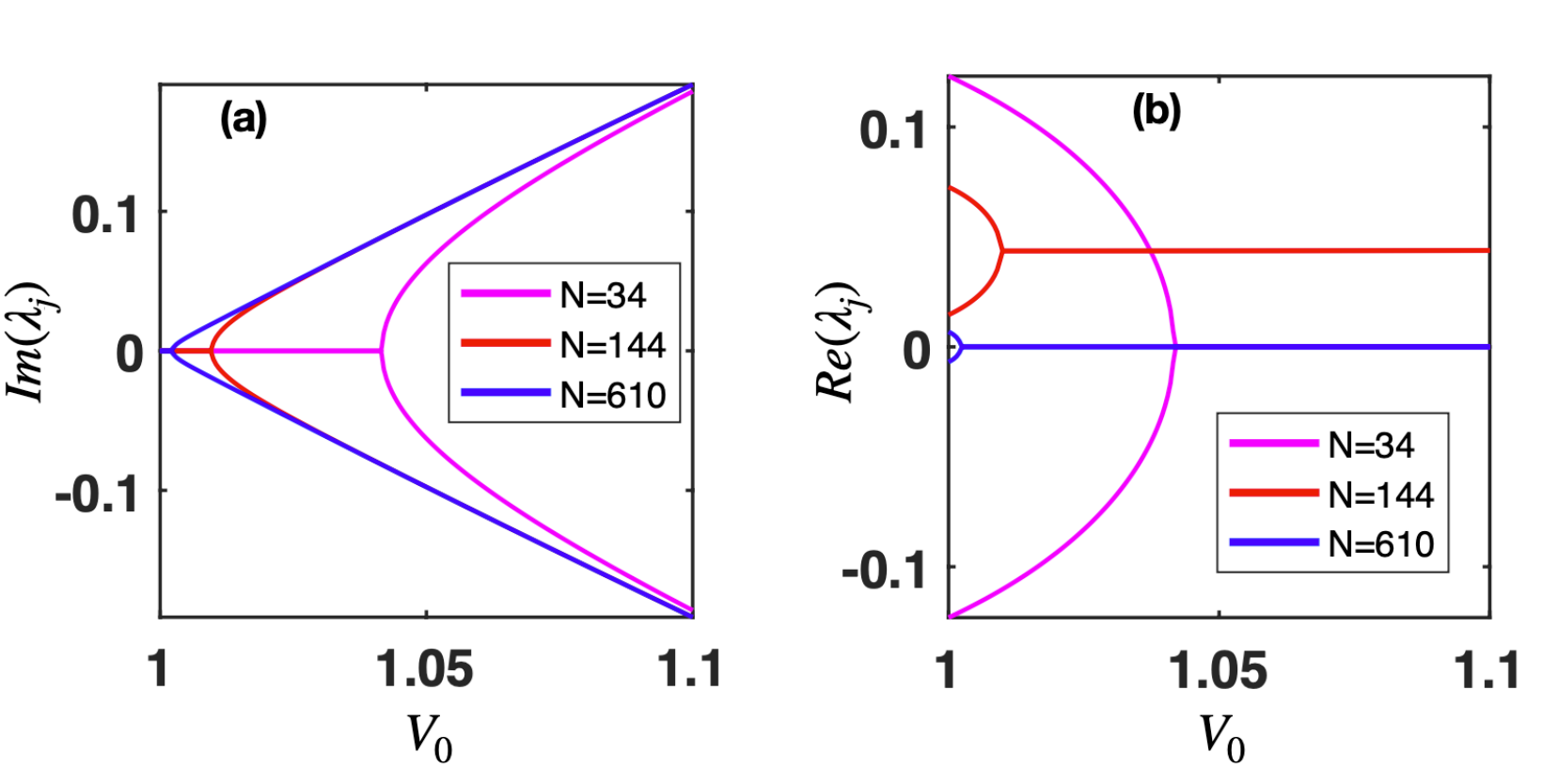}
\caption{(a) The maximum and minimum of imaginary parts of eigenvalues for Hamiltonian in Eq.~(\ref{H},\ref{V}) are plotted versus $V_0$ for different sizes (but always even) of the system with $J=1$. (b) The corresponding real parts of eigenvalues are plotted against $V_0$. The parameters used are $J=1$, $\alpha=21/34,89/144,377/610$ for $N=34,144$ and $610$ respectively. We find clear evidence that critical point approaches one as we increase $N$. }
	%\end{center}
  \label{app1}
\end{figure}

\begin{figure}%[h]
	%\begin{center}
\hspace*{-0.5 cm} \includegraphics[scale=.34]{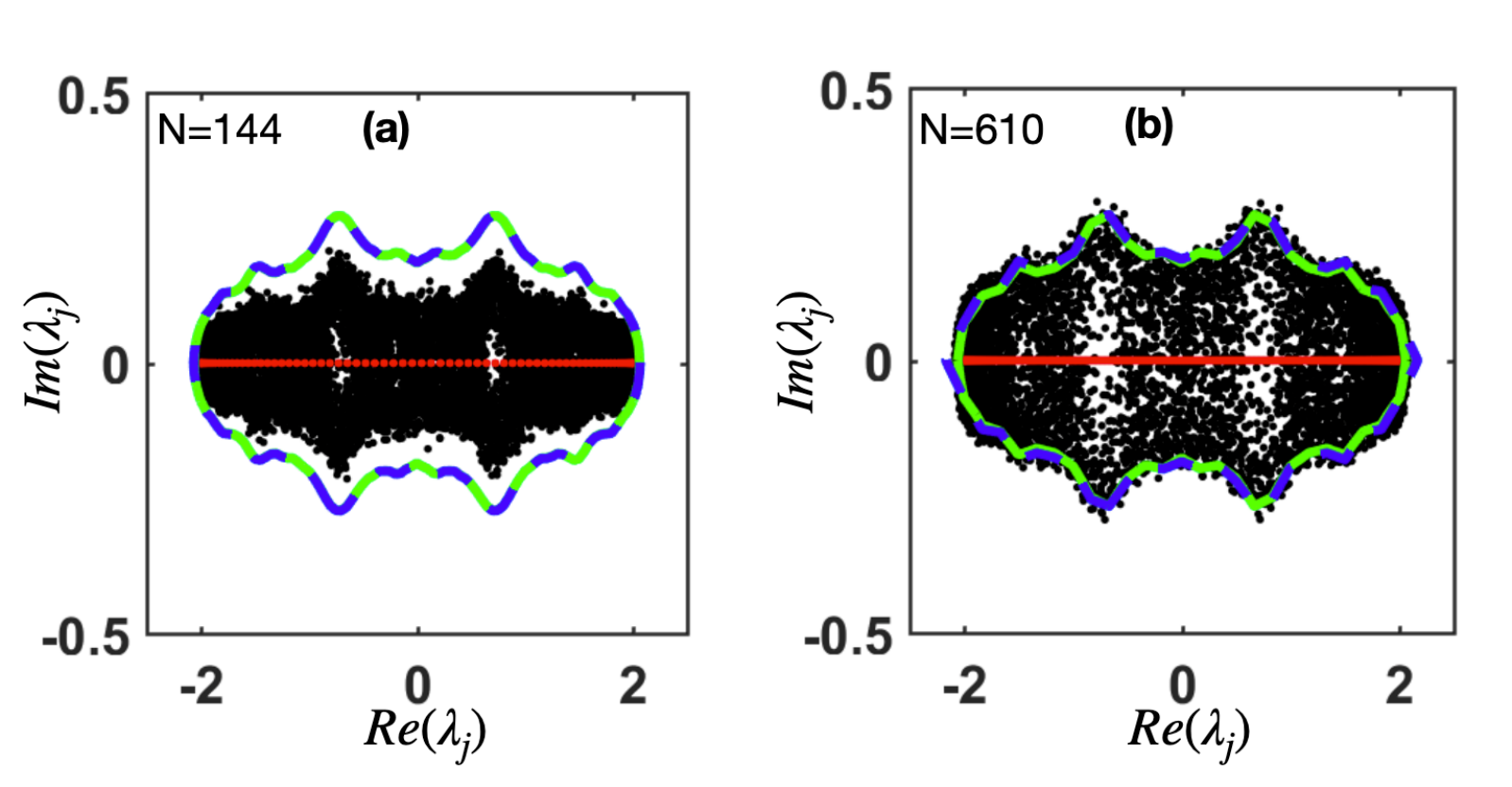}
\caption{(a) Comparison of the three pseudospectra methods discussed in Sec.~\ref{sec:sens} for $N=144$. The parameters used were $\alpha=89/144$ and $\epsilon=0.05$. We choose the value of $V_0$ to be at the transition point, i.e., $V_0=V_{0,C}$ . (b) Pseudospectra presented for $N=610$ from all three methods discussed in Sec.~\ref{sec:sens}. The parameters used were $\alpha=377/610$ and $\epsilon=0.05$ and $V_0$ was again chosen to be at the transition point $V_0=V_{0,C}$. The red dots show the spectrum in both figures. }
	%\end{center}
  \label{app2}
\end{figure}

\subsection{Value of the $\mathcal{PT}$-transition point for different lattice sizes}
\label{app:pt_tp}

The critical value for the $\mathcal{PT}$-phase transition point $V_{0,C}$ has a system size dependence.  
We showcase this in Fig.~\ref{app1}(a) where we plot the minimum and maximum of the imaginary part of eigenvalues with respect to $V_0$ for three even system sizes $N=34,144,610$ and $J=1$. Their corresponding real parts are plotted in Fig.~\ref{app1}(b). Pink, red and blue lines correspond to $N=34,144,610$ respectively. As the system size increases, the critical value of the phase transition approaches one. A few examples for the critical values are $V_{0,C}\sim 1.042$, $V_{0,C} \sim 1.01$ and $V_{0,C}\sim 1.002$ for $N=34,144$ and $610$ respectively thereby confirming that the critical point approaches to one in the large-$N$ limit.

\subsection{Pseudospectra for different lattice sizes}
\label{sec:app_pseudo}

Here, we study the pseudospectra for larger system sizes. 
Fig.~\ref{app2} shows the consistency between the three pseudospectra methods described in the main text (Sec.~\ref{sec:sens}). The systems size in Fig.~\ref{app2}(a,b) is $N=144$ and $N=610$ respectively. The systems are at the their corresponding phase transition points $V_{0,C}/J$ and the perturbation strength is 
$\epsilon=0.05$. We find that as we increase $N$, the pseudospectra becomes independent of system size.

\section{The almost orthogonal approximation}
\label{sec:overlap}

\setcounter{figure}{0}

\begin{figure}%[h]
	%\begin{center}
\hspace*{-0.71 cm} 
\includegraphics[scale=.32]{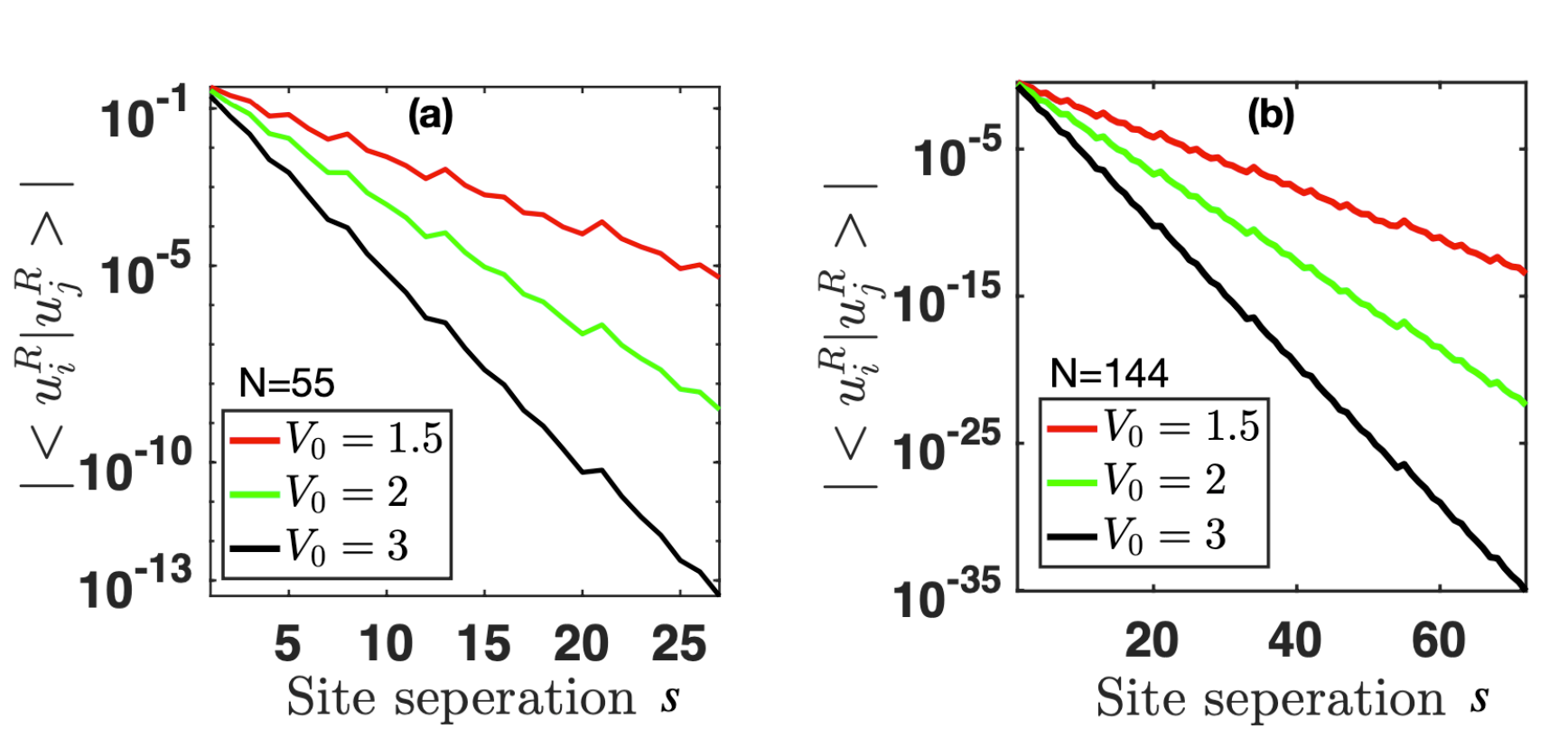}

\caption{Orthogonality $|\mathcal{O}_{i,j}|=|\langle u_{j}^R|u_{i}^R\rangle|$ in log-scale between the right eigenstates for the Hamiltonian in Eq.~(\ref{H},\ref{V}) as a function of site separation $s$ for various potential strengths $V_0$ with $J=1$. (a) $N=55, \alpha=34/55$  (b) $N=144,\alpha=89/144$.}
	%\end{center}
\label{app3}
\end{figure}

In this Appendix, we examine the degree of non-orthogonality and its implications to the dynamics. For non-Hermitian Hamiltonians, the right eigenstates are not generally orthogonal to each other. However, when eigenstates are localized they tend to have naturally small overlap with eigenstates localized far away from them. When the eigenstates are very localized they are almost orthogonal to all other eigenstates [see Eq.~\ref{Gamma}]. In other words, each eigenstate (say $i^{th}$) has most of its weight on a given site index (say $m_i$). This approximation is key to semi-analytically calculate the jump positions discussed in Sec.~\ref{sec:diff}. This methodology is referred to as the almost orthogonal approximation~\cite{jumps3}. To check the validity of the almost orthogonal approximation, we plot the average overlap of the eigenstates $\mathcal{O}_{i,j}=\langle u_{j}^R|u_{i}^R\rangle$ in  Fig.~\ref{app3} as a function of their site separation. By site separation, we mean the difference between the site indices at which each eigenstate ($i^{th}$ and $j^{th}$) is localized. If the $i^{th}$ eigenstate is localized at the site $m_i$, the site separation between eigenstates $|u_{i}^R\rangle$ and $|u_{j}^R\rangle$ is $s= |m_i-m_j|$. Our computations reveal two points. Firstly the overlap $\mathcal{O}_{i,j}=\langle u_{j}^R|u_{i}^R\rangle$ between eigenstates localized around distant sites falls off exponentially with site separation $s$. Secondly, as we increase $V_0$, the overlap between eigenstates decreases. Therefore, the almost orthogonal approximation [Eq.~\ref{Gamma}] discussed in Sec.~\ref{sec:diff} is justified and it becomes even better as $V_0$ increases. We find that these observations hold for any $N$. This is elucidated by the similarity between Fig.~\ref{app3}(a) and Fig.~\ref{app3}(b) which are for system sizes $N=55$ and $N=144$ respectively.

\end{document}